\documentclass{llncs}
\usepackage[british]{babel}

\usepackage{graphicx}
\usepackage{latexsym}

\usepackage{pgfplots}

\usepackage{stmaryrd} 
\usepackage{amssymb} 
\usepackage{amsmath}
\usepackage{refcount}

\usepackage{tikz}
\usetikzlibrary{arrows,automata,decorations.pathmorphing,backgrounds,positioning,fit,trees}

\newcommand{\shiq}{\ensuremath{\mathcal{SHIQ}}}
\newcommand{\shi}{\ensuremath{\mathcal{SHI}}}

\newcommand{\T}{\ensuremath{\mathcal{T}}}
\newcommand{\I}{\ensuremath{\mathcal{I}}}
\newcommand{\II}{\ensuremath{\mathcal{I'}}}
\newcommand{\sym}{\ensuremath{\sigma}}
\newcommand{\A}{\ensuremath{\mathcal{A}}}

\newcommand{\false}{\ensuremath{\mathit{false}}}

\newcommand{\K}{\ensuremath{\mathcal{K}}}

\newcommand{\Neg}{\ensuremath{\mathit{Neg}}}
\newcommand{\ABox}{\ensuremath{\mathit{ABox}}}

\newcommand{\DL}{\ensuremath{\mathit{DL}}}
\newcommand{\Negfalse}{\ensuremath{\mathit{Negfalse}}}
\begin{document}
\tikzstyle{every path}=[line width=1pt]
\title{Semantically Guided Evolution of $\mathcal{SHI}$ ABoxes}

\author{Ulrich Furbach \and Claudia Schon  \institute{University of Koblenz-Landau, Germany, email:\{uli,schon\}@uni-koblenz.de}}

\maketitle
\bibliographystyle{plain}

\begin{abstract}This paper presents a method for the evolution of \shi{} ABoxes which is based on a compilation technique of the knowledge base. For this the ABox is regarded as an interpretation of the TBox which is close to a model. It is shown, that the ABox can be used for a semantically guided transformation resulting in an equisatisfiable knowledge base. We use the result of this transformation to efficiently delete assertions from the ABox. Furthermore, insertion of assertions as well as repair of inconsistent ABoxes is addressed. For the computation of the necessary actions for deletion, insertion and repair, the E-KRHyper theorem prover is used.
\end{abstract}

\section{Introduction}

Description Logic knowledge bases consist of two parts: the TBox and the ABox. The TBox contains the terminological knowledge and describes the world using so called concepts and roles. The ABox contains knowledge about individuals, stating to which concepts they belong to and via which roles they are connected.
There is a considerable amount of work introducing update algorithms and 
 mechanisms for Description Logic knowledge bases, which  is of great interest to the Semantic Web community (see \cite{Halashek-Wiener:2006:DLR:2149716.2149777,LiLuMiWo-AIJ11} for details). It is an indisputable fact, that in practice, knowledge bases are subject to frequent changes (\cite{KR-2012-evolution}) and that even the construction of a knowledge base can be seen as an iterative process.
On the other hand this abets inconsistencies in knowledge bases. Therefore the removal of inconsistencies from knowledge bases is of great interest as well (\cite{DBLP:conf/sum/HorridgePS09}).
In this paper we are interested in an evolution of the knowledge base on the instance level. For this, we consider the TBox to be fixed and consistent. We address three different operations on the instance level of the knowledge base: \emph{deletion}, \emph{insertion} and \emph{repair}. Instance-level deletion means the deletion of an instance assertion from the deductive closure or the knowledge base by removing as few assertions as possible. Instance-level insertion means adding an instance assertion to the knowledge base. In both cases it is important that the resulting knowledge base is consistent.  For the task of ABox repair we are given an inconsistent knowledge base with consistent terminological part. The aim is to remove assertions from the ABox such that the resulting ABox together with the TBox is consistent. In all three tasks the changes performed should be minimal. 
This corresponds to the goal of maintaining as much from the original ABox as possible. This view of minimal change corresponds to a \emph{formula based} approach as opposed to a \emph{model based} approach as investigated in \cite{LiLuMiWo-AIJ11}. In the model based approach the set of models of the knowledge base resulting form a change operation should be as close as possible to the set of models of the original knowledge base.

%

In \cite{DBLP:conf/dlog/LenzeriniS11}, \cite{DeGiacomo2009IUE16669451666953} and  \cite{DBLP:conf/amw/CalvaneseKNZ10} instance level deletion, insertion and repair are addressed for DL-Lite knowledge bases.  In \cite{DBLP:conf/rr/LemboLRRS10} inconsistent DL-Lite ABoxes are considered. \cite{DBLP:conf/rr/LemboLRRS10} establishes inconsistency-tolerant semantics in order to be able to use those inconsistent ABoxes for query answering. \cite{Rosati:2011:CDI:2283516.2283574} studies the complexity of reasoning under inconsistent-tolerant semantics. Algorithms for the calculation of minimal repair of DL-Lite ABoxes suggested in  \cite{DBLP:conf/rr/LemboLRRS10} test the satisfiability of every single ABox assertion and every pair of ABox assertions w.r.t. the TBox. Since for DL-Lite the satisfiability test is tractable, this approach is reasonable. However the ExpTime completeness of  consistency testing of \shi{} ABoxes forbids such an approach. Further the algorithms suggested in  \cite{DBLP:conf/rr/LemboLRRS10} cannot be used for \shi{} ABoxes, because these algorithms exploit the following nice property of DL-Lite: as shown in \cite{DBLP:conf/amw/CalvaneseKNZ10}, in DL-Lite the unsatisfiability of an ABox w.r.t. a TBox is either caused by a single assertion or a pair of assertions. However in \shi{} an arbitrary number of assertions can cause unsatisfiability w.r.t. a TBox.

Our approach is motivated by the observation that a consistent ABox can be seen as a (partial) model of the TBox, which can be used to guide the reasoning process, as proposed in \cite{chpl:94a}. In  \cite{baumgartner97semantically} this approach was used 
for model-based diagnosis, where an initial interpretation, which is very close to a model, was used to compute the deviations of a minimal model to this interpretation. 
 In \cite{AravindanBaumgartner99} the  same approach was applied to  view deletion in databases.
In our case it is reasonable to assume, that the ABox is very close to a model of the TBox. 
We use this assumption to semantically guide the construction of instance-based deletion, insertion and repair of ABoxes. As in \cite{baumgartner97semantically}, we gradually revise the assumption of the given ABox being a model for the TBox. This leads to a natural construction of \emph{minimal} instance deletions/insertions and repairs of ABoxes.

The advantage of this approach is that there is no need to define new algorithms for updates and repair, which have to be proven correct. Instead we  will use a static compilation of the knowledge base according to the update or repair requirement. We prove that this transformation preserves the necessary semantics. A theorem prover can be used to compute the necessary update and repair actions. A hypertableau-based theorem prover like E-KRHyper is very well suited for this task, because the transformation enables it to calculate only the deviation of the ABox. 
Since E-KRHyper has recently been extended to deal with knowledge bases given in $\mathcal{SHIQ}$ \cite{cadesd}, we chose to use this theorem prover.

Our approach is related to axiom pinpointing. For a given consequence, axiom pinpointing is the task to find the minimal subsets of the knowledge base under consideration, having this consequence. See \cite{DBLP:journals/logcom/BaaderP10} for details. 
In \cite{horridge2008laconic} laconic and precise justifications are introduced. Given an ontology and an entailment, a justification is a minimal subset of that ontology such that the entailment still holds in the subset. Roughly spoken, laconic justifications are not allowed to contain superfluous parts. In contrast to axiom pinpointing and justifications, we calculate subsets of the ABox and not of the whole knowledge base.
In \cite{DBLP:conf/ijcai/SchlobachC03} incoherent TBoxes, i.e. TBoxes containing an unsatisfiable concept, are investigated.
\cite{Halashek-Wiener:2006:DLR:2149716.2149777} considers so called syntactic ABbox updates. Similar to our approach, assertions are added to or removed from the ABox. In contrast to our approach, it is neither guaranteed that the removed assertion is not contained in the deductive closure nor that the result of adding the assertion is consistent.

In Section \ref{sect:shi} we give both syntax and semantics of the Description Logic \shi{}. In addition to that, we introduce the notion of DL-clauses as used in \cite{msh07optimizing}. In Section \ref{sect:aboxevolution} we give definitions for instance-level deletion, insertion and repair. Section \ref{sect:nf} introduces the so called $\K{}^ {*}$-transformation which in Section \ref{sect:using} is used to calculate the  instance-level deletion, insertion and repair. The $\K{}^{*}$-transformation is implemented and in Section \ref{sect:experiments} we present first experimental results.
Proofs of all theorems, propositions and lemmas can be found in \cite{FurbachSchonTR2013}.

\section{\shi{} and DL-Clauses}\label{sect:shi}

First, we introduce the Description Logic \shi{}{}. 
Given a set of \emph{atomic roles} $N_R$, the set of \emph{roles} is defined as $N_R \cup \{ R^- \mid R \in N_R \}$, where $R^-$ denotes the \emph{inverse role} corresponding to the atomic role $R$. Let further $Inv$ be a function on the set of roles that computes the inverse of a role, with $Inv(R) = R^-$ and $Inv(R^-) = R$.
A \emph{role inclusion axiom} is an expression of the form $R \sqsubseteq S$, where $R$ and $S$ are atomic or inverse roles. A \emph{transitivity axiom} is of the form $\mathit{Trans}(S)$ for $S$ an atomic or inverse role. An RBox $\mathcal{R}$ is a finite set of role inclusion axioms and transitivity axioms.
$\sqsubseteq^*$ denotes the reflexive, transitive closure of $\sqsubseteq$ over $\{ R \sqsubseteq S, Inv(R) \sqsubseteq Inv(S) \mid R \sqsubseteq S \in \mathcal{R} \}$. A role $R$ is \emph{transitive} in $\mathcal{R}$ if there exists a role $S$ such that $S \sqsubseteq^* R$, $R \sqsubseteq^* S$, and either $\mathit{Trans}(S) \in \mathcal{R}$ or $\mathit{Trans}(Inv(S)) \in \mathcal{R}$. If no transitive role $S$ with $S \sqsubseteq^* R$ exists, $R$ is called \emph{simple}.

Let $N_C$ be the set of \emph{atomic concepts}. The set of \emph{concepts} is then defined as the smallest set containing $\top$, $\bot$, $A$, $\lnot C$,  $C \sqcap D$, $C \sqcup D$, $\exists R.C$ and $\forall R.C$ for $A \in N_C$, $C$ and $D$ concepts and $R$ a role.

A \emph{general concept inclusion} (GCI) is of the form $C \sqsubseteq D$, and a TBox $\mathcal{T}$ is a finite set of GCIs.

Given a set of individuals $N_{I}$, an ABox $\A{}$ is a finite set of assertions of the form $A(a)$ and $R(a,b)$, with $A$ an atomic concept, $R$ an atomic role and $a$, $b$ individuals from $N_{I}$.  Note that in our setting, the ABox is only allowed to contain assertions about the belonging of individuals to atomic concepts and roles.

A knowledge base $\mathcal{K}$ is a triple $\mathcal{(R, T, A)}$ with signature $\Sigma=(N_{C},N_{R},N_{I})$.
The tuple $\I{} = (\cdotp^\I{}, \Delta^\I{})$ is an \emph{interpretation} for $\mathcal{K}$ iff $\Delta^\I{}$ is a nonempty set and $\cdotp^\I{}$ assigns an element $a^\I{} \in \Delta^\I{}$ to each individual $a$, a set $A^\I{} \subseteq \Delta^\I{}$ to each atomic concept $A$, and a relation $R^\I{} \subseteq \Delta^\I{} \times \Delta^\I{}$ to each atomic role $R$. $\cdotp^\I{}$ then assigns values to more complex concepts and roles as described in Table \ref{tab:dl-shiq-semantics}. $\I{}$ is a \emph{model} of $\mathcal{K}$ ($\I{} \models \mathcal{K}$) if it satisfies all axioms and assertions in $\mathcal{R}$, $\mathcal{T}$ and $\mathcal{A}$ as shown in Table \ref{tab:dl-shiq-semantics}. A TBox $\mathcal{T}$ is called consistent, if there is an interpretation satisfying all axioms in $\mathcal{T}$.
A concept $C$ is called \emph{satisfiable w.r.t.} $\mathcal{R}$ and $\mathcal{T}$ iff there exists a model $\I{}$ of $\mathcal{R}$ and $\mathcal{T}$ with $C^\I{} \neq \emptyset$. 

\begin{table}[ht]
\begin{center}
\begin{tabular}{r c l r c l}
\hline
\multicolumn{6}{c}{Concepts and Roles} \\
\hline \\[-3mm]
$\top^\I{}$ & = & $\Delta^\I{}$ & $(R^-)^\I{}$ & = & $\{ (y, x) \mid (x, y) \in R^\I{} \}$ \\
$\bot^\I{}$ & = & $\emptyset$ & $(\forall R.C)^\I{}$ & = & $\{ x \mid \forall y: (x, y) \in R^\I{} \Rightarrow y \in C^\I{}$ \\
$(\neg C)^\I{}$ & = & $\Delta^\I{} \backslash C^\I{}$ & $(\exists R.C)^\I{}$ & = & $\{ x \mid \exists y: (x, y) \in R^\I{} \wedge y \in C^\I{} \}$ \\
$(C \sqcup D)^\I{}$ & = & $C^\I{} \cup D^\I{}$ &\hspace{3cm} \\
$(C \sqcap D)^\I{}$ & = & $C^\I{} \cap D^\I{}$ &\\
\end{tabular}
\begin{tabular}{p{1.15cm} r c l p{1cm} r c l p{1.15cm}}
\hline
& \multicolumn{3}{c}{TBox \& RBox axioms} & & \multicolumn{3}{c}{ABox axioms} & \\
\hline \\[-3mm]
& $C \sqsubseteq D$ & $\Rightarrow$ & $C^\I{} \subseteq D^\I{}$ & & $C(a)$ &  $\Rightarrow$ & $a^\I{} \in C^\I{}$ \\
& $R \sqsubseteq S$ & $\Rightarrow$ & $R^\I{} \subseteq S^\I{}$ & & $R(a, b)$ & $\Rightarrow$ & $(a^\I{}, b^\I{}) \in R^\I{}$ \\
& $\mathit{Trans}(R)$ & $\Rightarrow$ & $(R^\I{})^+ \subseteq R^\I{}$ & & \\
\end{tabular}
\end{center}
\caption{Model-theoretic semantics of \shi{}. $R^+$ is the transitive closure of $R$.}
\label{tab:dl-shiq-semantics}
\end{table}

In the sequel we adapt the notion of DL-clauses introduced in \cite{msh07optimizing} to the Description Logic \shi{}. These DL-clauses allow to use existent theorem provers which are based on the hypertableau calculus to compute models or to decide satisfiability. 
DL-clauses are universally quantified implications of the form $\bigvee V_{j} \leftarrow \bigwedge U_{i}$:
\begin{definition}(\cite{msh07optimizing}) An atom is of the form $B(s)$, $R(s,t)$, $\exists R.B(s)$ or $\exists R.\lnot B(s)$  for $B$ an atomic concept and $s$ and $t$ individuals or variables. An atom not containing any variables is called a ground atom. A DL-clause is of the form $V_{1} \lor \ldots \lor V_{n} \leftarrow U_{1} \land \ldots \land U_{m}$ with $V_{i}$ atoms and $U_{j}$ atoms of the form $B(s)$ or $R(s,t)$ and $m \geq 0$ and $n \geq 0$. If $n=0$, we denote the left hand side (head) of the DL-clause by $\bot$. If $m=0$, we denote the right hand side (body) of the DL-clause by $\top$.
\end{definition}

\begin{definition}(Semantics of DL-clauses; \cite{msh07optimizing}) \label{def:stransformeddlclauses} Let $V_{1} \lor \ldots \lor V_{n} \leftarrow U_{1} \land \ldots \land U_{m}$  be a DL-clause and $N_{V}$ a set of variables, disjoint from $N_{I}$.
Let further $\mathcal{I}=(\Delta^{\mathcal{I}},\cdot^{\mathcal{I}})$ be an interpretation and $\mu:N_{V}\rightarrow \Delta^{\mathcal{I}}$ be a variable mapping.
Let $a^{\mathcal{I},\mu} = a^{\mathcal{I}}$ for an individual $a$ and $x^{\mathcal{I},\mu} = \mu(x)$ for a variable $x$. 
Satisfaction of an atom, a DL-clause, and set of DL-clauses $N$ in $\mathcal{I}$ and $\mu$ is defined as follows:

\begin{tabular}{l l}
$\mathcal{I},\mu \models C(s)$ & if $s^{\mathcal{I},\mu} \in C^{\mathcal{I}}$\\
$\mathcal{I},\mu \models R(s,t)$ & if $\langle s^{\mathcal{I},\mu}, t^{\mathcal{I},\mu} \rangle \in R^{\mathcal{I}}$\\
$\mathcal{I},\mu \models \bigvee\limits_{j=1}^{n} V_{j} \leftarrow \bigwedge\limits_{i=1}^{m} U_{i}$ & if $\mathcal{I},\mu \models V_{j}$ for some $1 \leq j \leq n$ whenever $\mathcal{I},\mu \models U_{i}$\\
&for each $1 \leq i \leq m$ \\
$\mathcal{I}\models \bigvee\limits_{j=1}^{n} V_{j} \leftarrow \bigwedge\limits_{i=1}^{m} U_{i}$ & if $\mathcal{I},\mu \models \bigvee_{j=1}^{n} V_{j} \leftarrow \bigwedge_{i=1}^{m} U_{i}$ for  all mappings $\mu$\\
$\mathcal{I}\models N$ & if $\mathcal{I} \models r$ for each DL-clause $r \in N$
\end{tabular}
\end{definition}

We will not give the transformation into DL-clauses. The details can be found in \cite{msh07optimizing}.  The transformation avoids an exponential blowup by using the well-known structural transformation \cite{DBLP:journals/jsc/PlaistedG86} and can be computed in polynomial time.

By  $\Xi(\T)$ ($\Xi(\A)$) we denote the set of DL-clauses for a TBox $\T{}$ (an ABox $\A{}$). For a knowledge base $\mathcal{K}=(\T,\A)$,  $\Xi(\mathcal{K})=\Xi(\T) \cup \Xi(\A)$. 
According to \cite{msh07optimizing}  for every interpretation $\I{}$, $\I \models \mathcal{K}$ iff $\I \models \Xi(\T)$ and $\I \models \A{}$. 

Since we assume the ABox assertions to be atomic, the ABox itself corresponds to a set of DL-clauses.

\begin{example}\label{ex:dlclauses} The TBox $\mathcal{T}=\lbrace B\sqsubseteq \exists R.C, \exists R.C\sqsubseteq D, D\sqsubseteq C \rbrace$ corresponds to the set of DL-clauses $\Xi(\T)=\lbrace 
\exists R.C(x) \leftarrow B(x), D(x) \leftarrow R(x,y) \land C(y), C(x) \leftarrow D(x)\rbrace$.
\end{example}

Sometimes it is convenient to regard both the body and the head of a DL-clause $C$ as a set of atoms like $C = \mathbf{H}\leftarrow \mathbf{B}$. This allows us to write $A \in \mathbf{B}$  ($A \in \mathbf{H}$) if atom $A$ occurs in the body (in the head) of DL-clause $C$.
The signature of a set of DL-clauses is the set of atomic concepts and atomic roles occurring in the DL-clause. The size of a DL-clause $C$ is defined as the numbers of atoms occurring in $C$ and is denoted $\mathit{size}(C)$. The size of a set of DL-clauses $N$ denoted by $\mathit{size}(N)$ is the sum of sizes of all DL-clauses in $N$.
In the sequel we need a function extracting the concept/role from an atom:
\begin{definition}(Symbol Extraction Function)
Let $A$ be an atom. Then $\sym(A)$ is defined as follows:\\
 $
 \sym(A) = \begin{cases} 
 B	  		& \mbox{if}\ A=B(s) \mbox{ for some atomic concept } B,\\
 R	  		& \mbox{if}\ A=R(s,t)  \mbox{ for some atomic role } R,\\
 \exists R.B  	& \mbox{if}\  A=\exists R.B(s)  \mbox{ for some atomic role } R \mbox{ and}\\
			& B=E \mbox { or } \lnot E \mbox{ for some atomic concept } E.
 \end{cases} 
$
\end{definition}
By $\sym(N)$ for a set of atoms $N$ we denote the union of $\sym(A)$ for all atoms $A \in N$.

In the following it is convenient for us to regard an interpretation as the set of ground atoms assigned to true by the interpretation. A set of ground atoms and an interpretation can be seen as equivalent, since every set of ground atoms uniquely determines a Herbrand interpretation.
Now we can introduce the idea of minimal models to DL-clauses.  

\begin{definition}(Minimal Model for a Set of DL-Clauses)
Let $\DL$ be a set of DL-clauses. An Interpretation $\I{}$ is called a minimal model for $\DL$, iff $\I{}$ is a model for $\DL$ and further there is no model $\II$ for $\DL$ such that $\II \subset \I{}$.
 \end{definition}
 Next we define minimality of models w.r.t. a set of ground atoms. We will later use this notion in order to minimize the number of ABox assertions which are to be deleted.
 \begin{definition}($\Gamma$ Minimal Model)
Let $\DL$ be a set of DL-clauses and  $\Gamma$ be a set of ground atoms. An interpretation $\I{}$ is a $\Gamma$-minimal model for $\DL$ iff $\I{}$ is a model for $\DL$ and further there is no model $\II$ for $\DL$ with  $\II \cap \Gamma \subset \I \cap \Gamma$.
 \end{definition}

\section{ABox Evolution}\label{sect:aboxevolution}
We address three different operations on the instance level of the knowledge base: deletion, insertion and repair. 
The first scenario we are considering is the following: given a knowledge base $\mathcal{K}=(\T,\A)$ with a consistent TBox $\T{}$, we want to remove an ABox assertion for example $A(a)$ from the ABox. In general it is not sufficient to only delete $A(a)$ from the ABox, because $A(a)$ can still be contained in the deductive closure. So the task is to determine a minimal set of ABox assertions, which have to be deleted from the ABox in order to prevent that $A(a)$ is a logical consequence of the knowledge base. 
This leads to the following definition.
\begin{definition}(Minimal Instance Deletion) Let $\mathcal{K}=(\T,\A)$ be a knowledge base where $\T{}$ is consistent. A ground atom $D$ of the form $A(a)$ or $R(a,b)$ with $D \in \A{}$ is called delete request. Further $\A' \subseteq \A{}$   is called minimal instance deletion of $D$ from $\A{}$ if $\T{}  \cup \A' \not\models D$ and there is no $\A'' $ with $\A' \subset \A'' \subseteq \A{}$ and $\T{}  \cup \A'' \not\models D$.  
\end{definition}
\begin{example}\label{ex:minimalinstancedeletion} We consider the ABox
\begin{equation*} 
\mathcal{A} = \lbrace B(a), D(a), C(b), R(b,b), R(a,a) \rbrace
\end{equation*} 
 together with the TBox given in Example \ref{ex:dlclauses}.
The delete request $D(a)$ has a minimal instance deletion 
 \begin{equation*}
 \A'=\lbrace C(b), R(b,b), R(a,a) \rbrace
 \end{equation*}
\end{example}

Next we want to repair an ABox which is not consistent w.r.t. its TBox. 
\begin{definition}(Minimal ABox Repair) Let $\mathcal{K}=(\T,\A)$ be a knowledge base where $\T{}$ is consistent.  $\A' \subseteq \A{}$   is called minimal ABox repair of $\A{}$ if $\T{}  \cup \A'$ is consistent and there is no $\A'' $ with $\A' \subset \A'' \subseteq \A{}$ and $\T{}  \cup \A''$ consistent.
\end{definition}
Note that we define the notion of a minimal ABox repair in a way, that it is also applicable to an ABox which is consistent to its TBox. In this case, the minimal ABox repair corresponds to the original ABox.

The third instance level operation we address is insertion of an assertion into an existing ABox. The problem that arises when considering insertion is, that the resulting ABox might be inconsistent w.r.t. its TBox. 
\begin{example}\label{ex:insertion}Let us consider the set of DL-clauses  
\begin{equation*} 
\Xi(\T) =\lbrace \bot \leftarrow C(x) \land D(x)\rbrace
\end{equation*} 
together with the ABox 
\begin{equation*} 
\A=\lbrace C(a) \rbrace
\end{equation*} 
Adding the assertion $D(a)$ into $\A{}$ leads to $\A'=\lbrace C(a), D(a) \rbrace$, which is inconsistent w.r.t. $\T{}$. 
\end{example}
We avoid inconsistent results by the next definition.

\begin{definition}(Minimal Instance Insertion) Let $\mathcal{K}=(\T,\A)$ be a knowledge base with $\T{}$ consistent and $D$ a ground atom of the form $A(a)$ or $R(a,b)$. An ABox $\A'$ is called minimal instance insertion of $D$ into $\A{}$ if 
\begin{itemize}
\item $D \in \A'$, 
\item $(\A'\setminus D) \subseteq \A{}$,   
\item $\T{}  \cup \A'$ is consistent and there is no $\A'' $ with $D \in \A''$ and $(\A'\setminus D) \subset (\A''\setminus D) \subseteq \A{}$ and $\T{}  \cup \A''$ is consistent.  
\end{itemize}
\end{definition}

\section{$\mathcal{K}^{*}$-Transformation}\label{sect:nf}
We will solve the tasks defined in Section \ref{sect:aboxevolution} by using the $\K{}^ {*}$-transformation which will be introduced in this section. As discussed in the introduction we want to use the ABox of the knowledge base as a partial model, which will guide our transformation.

Considering the task of deleting a given instance, we want to determine a minimal set of ABox assertions which have to be deleted in order to prevent the instance from being contained in the deductive closure of the knowledge base. 
The idea of the transformation we are about to use was introduced in \cite{baumgartner97semantically}. We replace occurrences of an atom $A(a)$ in a clause by $\lnot \mathit{NegA}(a)$. This transformation can be seen as switching the sides in the clause representation of DL-clauses. This makes sense, when a bottom-up proof procedure like E-KRHyper is used: a fact $A(a)\leftarrow $ changes the side and the clause becomes $\leftarrow \mathit{NegA}(a)$. As a consequence $A(a)$ is not derived explicitly. It is assumed to be in the model until the opposite has to be derived. 

Deducing an atom $\mathit{NegA}(a)$ means that we have to revise the ABox and that we have to remove atom $A(a)$ from the ABox.  By using this transformation we only need to calculate the atoms we have to remove from the ABox. All remaining atoms will be kept in the ABox. Since it is reasonable to expect the ABox to be very large, it is advantageous to calculate only the deviation from the original ABox.


\begin{definition} The $\Neg$ and the $\ABox{}$ function map atoms to renamed atoms:
\begin{itemize}
\item For atomic concepts $A$ and an individual or variable $a$: 
	\begin{itemize}
		\item $\Neg(A(a))=\mathit{NegA}(a)$
		\item $\ABox{}(A(a))=\mathit{ABoxA}(a)$
	\end{itemize}
\item For atomic roles $R$ and individuals or variables $a$, $b$:
	\begin{itemize}
		\item $\Neg(R(a,b))=\mathit{NegR}(a,b)$
		\item $\ABox{}(R(a,b))=\mathit{ABoxR}(a,b)$
	\end{itemize}
\end{itemize}
\end{definition}
We slightly abuse notation by using the $\Neg$ function to rename atomic concepts and atomic roles: for $B$ an atomic concept or an atomic role: $\Neg(B)=\mathit{NegB}$.
Further for a set of atoms $P$, $\Neg(P)$ is defined as: $\Neg(P) = \lbrace \Neg(A)\ \vert \ A \in P\rbrace$.
So we can use the $\Neg$ function to rename atoms, sets of atoms and atomic concepts and roles.
\begin{definition}\footnote{Due to the helpful remarks of an anonymous reviewer of the DL Workshop, this definition was revised. These changes also affect the results presented in the experiments.} \label{def:renaming}(Renaming) Let $\DL$ be a set of DL-clauses and $S$ a set of atomic concepts and atomic roles. Let $C \in \DL$ be $C =  \mathbf{H} \leftarrow \mathbf{B}$. Then $\textsf{R}_{S}(C)$, the \emph{renaming of } $C$ w.r.t. $S$ is 
\begin{align}
&\textsf{R}_{S}(C) = 	\nonumber \\
	  	& \lbrace C \rbrace  \label{def:renam0}\\
&\cup \nonumber \\
	&\lbrace (\bigvee\limits_{\stackrel{A \in \mathbf{H},}{\sym(A) \notin S}}A) \lor (\bigvee\limits_{\stackrel{B \in \mathbf{B},}{\sym(B) \in S}}\hspace{-3mm}\Neg(B))  \leftarrow (\bigwedge\limits_{\stackrel{B \in \mathbf{B},}{\sym(B) \notin S}}B) \land  (\bigwedge\limits_{\stackrel{A \in \mathbf{H},}{\sym(A) \in S}} \Neg(A)) \rbrace \label{def:renam1}\\
	&\cup \nonumber \\
			& \lbrace \bot \leftarrow R(x,y) \land \mathit{NegR}(x,y)\ \vert\ \exists A \in (\mathbf{H} \cup \mathbf{B}) \text{ with }  \sigma(A)=R \in S \text{ or } \exists A \in \mathbf{H} \nonumber \\
			&  \hspace{43.25mm}\text{ of the form } A =\exists R.C(z) \text{ and }  R \in S\rbrace\label{def:renam2} \\
			&\cup \nonumber \\
			& \lbrace \bot \leftarrow D(x) \land \mathit{NegD}(x)\ \vert\  \exists A \in (\mathbf{H} \cup \mathbf{B}) \text{ with }  \sigma(A)=D \in S \text{ or } \nonumber \\ 
			&\hspace{37.75mm} \exists A \in \mathbf{H} \text{ of the form } A =\exists R.D(z) \text{ and }  R \in S\rbrace\label{def:renam3}
\end{align}
For a set of DL-clauses $\DL$, the renaming $\textsf{R}_{S}(\DL)$ w.r.t. $S$ is defined as the union of the renaming of all its clauses.
\end{definition}
 Note that renaming is a bijective function on a set of DL-clauses.
 Further renaming can be performed in time linear to the size of the set of DL-clauses times the size of $S$.

The next proposition states the fact, that renaming preserves satisfiability. Furthermore given a model  for a set of DL-clauses $\DL$, it is possible to calculate a model for the renamed set of DL-clauses $\textsf{R}_{S}(\DL)$ and vice versa.

\begin{proposition}(Renaming Models)\label{prop:renamingmodels} Let $\DL$ be a set of DL-clauses, $S$ a set of atomic concepts and atomic roles and $\mathcal{I}$ an interpretation. Then $\mathcal{I} \models \DL$ iff $\mathcal{I}^{S} \models \textsf{R}_{S}(\DL)$, where $\mathcal{I}^{S}$ and $\mathcal{I}$ have the same domain and the same interpretation of individuals. In addition to that  the interpretation of all roles and concepts occurring in $\DL$ coincide. Further $(\Neg(B))^{\mathcal{I}^{S}}=\overline{B^{\I}}$ for all concepts names $B \in S$ 
and $(\Neg(R))^{\mathcal{I}^{S}}=\overline{R^{\I}}$ for all atomic roles $R \in S$. 
\end{proposition}

\begin{definition}\label{def:kstar}($\mathcal{K}$*-Transformation) Let $\mathcal{K}=(\T, \A)$ be a knowledge base (where $\T{}$ is consistent). Let $S$ be the set of atomic concepts and atomic roles occurring in $\A{}$. Then $\K{}^{*}$ is the clause set obtained by renaming $\Xi(\T)$ w.r.t. $S$ and adding the set of DL-clauses
$\lbrace \ABox{}(A) \leftarrow \top \mid \text{for all assertions } A \in \A{} \rbrace$.
\end{definition}
We have to add $\lbrace \ABox{}(A) \leftarrow \top \mid \text{for all assertions } A \in \A{} \rbrace$ to the result of renaming for two reasons: first of all we have to  introduce the individuals occurring in the ABox to the theorem prover. Furthermore it is helpful to calculate minimal deletions.
\begin{proposition}\label{prop:equisat}Let $\mathcal{K}=(\T, \A)$ be a knowledge base and $S$ be the set of atomic concepts and atomic roles occurring in $\A{}$ and $\T{}$. Then $\Xi(\T)$, $R_{S}(\Xi(\T))$ and $\K{}^ {*}$ are equisatisfiable.
\end{proposition}

 \begin{example}\label{ex:*transformation} We consider the set of DL-clauses given in Example \ref{ex:dlclauses} together with the ABox $\mathcal{A} = \lbrace B(a), D(a), C(b), R(b,b), R(a,a) \rbrace$. Then $S=\lbrace B, D, C, R \rbrace$. Renaming the DL-clauses given in Example \ref{ex:dlclauses} w.r.t. $S$ leads to $\mathcal{K}^{*}$:
\begin{align*}
\exists R.C(x) &\leftarrow B(x). \\
\exists R.C(x) \lor \mathit{Neg }B(x) & \leftarrow  \top. \\
D(x)& \leftarrow R(x,y) \land C(y). \\
\mathit{Neg}R(x,y) \lor \mathit{Neg}C(y)& \leftarrow \mathit{Neg}D(x).\\
C(x) &\leftarrow D(x).\\
\mathit{Neg}D(x)& \leftarrow \mathit{Neg}C(x).\\
 \bot &\leftarrow R(x,y) \land \mathit{NegR}(x,y).\\
 \bot &\leftarrow C(x) \land \mathit{NegC}(x).\\
 \bot &\leftarrow B(x) \land \mathit{NegB}(x).\\
 \bot &\leftarrow D(x) \land \mathit{NegD}(x).\\
\ABox{} B(a) &\leftarrow \top.\\
\ABox{} D(a) &\leftarrow \top.\\
\ABox{} C(b) &\leftarrow \top.\\
\ABox{} R(b,b) &\leftarrow \top.\\
\ABox{} R(a,a) &\leftarrow \top.
\end{align*}
\end{example}
In the worst case the $\mathcal{K}^{\ast}$-transformation quadruples the size of a set of DL-clauses: $S$ is the set of all concepts/roles occurring in the clause set. The ABox contains $b$ assertions and the TBox consists of a single clause: 
$C = H_{1}\lor \ldots \lor H_{i} \leftarrow  B_{1}\land \ldots \land B_{j}$
with $n=i+j$. W.l.o.g. the symbols of all atoms occurring in $C$ are concepts.
This set of DL-clauses has the size $n+b$. Renaming results in:
\begin{align*}
\lbrace H_{1}\lor \ldots \lor H_{i}& \leftarrow  B_{1}\land \ldots \land B_{j},\\
\Neg(B_{1}) \lor \ldots \lor \Neg(B_{j}) &\leftarrow  \Neg(H_{1}) \land \ldots  \land  \Neg(H_{i}), \\
 \bot &\leftarrow \sigma(H_{1})(x)  \land \mathit{Neg}(\sigma(H_{1}))(x),\\
&\vdots\\
 \bot &\leftarrow \sigma(H_{i})(x)  \land \mathit{Neg}(\sigma(H_{i}))(x),\\
 \bot &\leftarrow \sigma(B_{1})(x)  \land \mathit{Neg}(\sigma(B_{1}))(x),\\
& \vdots \\
 \bot &\leftarrow \sigma(B_{j})(x)  \land \mathit{Neg}(\sigma(B_{j}))(x),\\
\cup &\lbrace  \ABox{}(A) \leftarrow \top \mid \text{for all assertions } A \in \A{} \rbrace
\end{align*}
The first clause is the original clause from the TBox. Its size is $n$.
The second clause is created by renaming and has size $n$. Then $n$ clauses of size $2$ follow. At the end of the clause set are $b$ clauses of the form $\ABox{}(A)$ each of size $1$. All in all the resulting set of clauses has the size $n + n + 2*n + b \leq 4 * (n+b)$, which is four times higher than the size of the original set of DL-clauses.

\section{Using the $\K{}^{*}$-transformation for ABox Evolution}\label{sect:using}
Firstly we address deletion:
Recall that according to the definition of the $\Neg$ function,  $\Neg(\A)$ is defined as $\lbrace \Neg(A) \ \vert\ A \in \A{} \rbrace$.
Next we show how to use $\Neg(\A)$-minimal models to calculate minimal instance deletions. For a given model $M$ we construct $Del(M)=\lbrace A \in \A\ \vert\ \Neg(A) \in M \rbrace$. Intuitively $Del(M)$ constitutes the set of  ABox assertions supposed to be deleted from the ABox to obtain a minimal instance deletion.

\begin{theorem}\label{theorem:minmodel}Let $\mathcal{K}=(\T,\A)$ be a knowledge base where $\T{}$ is consistent, $S$ the set of atomic concepts and atomic roles occurring in $\A{}$, and $D$ a delete request. Let $M^{S}$ be a $\Neg(\A)$-minimal model for $\K{}^ {*} \cup \lbrace \Neg(D)\leftarrow\top \rbrace$. Then $\A{} \setminus Del(M^{S})$ is a minimal instance deletion of $D$ from $\A{}$.
\end{theorem}
Proof by first showing $\T{}  \cup (\A{} \setminus Del(M^{S})) \not\models D$ by constructing a model according to Proposition \ref{prop:renamingmodels} for $\T{}  \cup (\A{} \setminus Del(M^{S})) \cup \lbrace \leftarrow D \rbrace$ from $M^{S}$. And then showing that there is no $Del'\subset Del(M^{S})$ with $\T{}  \cup (\A{} \setminus Del') \not\models D$. See \cite{FurbachSchonTR2013} for details.

\begin{example}Now we delete $D(a)$ from the DL-clauses of our running example. For this, we add the clause $\mathit{NegD(a)\leftarrow}$ to the result of the $\K{}^ {*}$ transformation given in Example \ref{ex:*transformation}.  For lack of space we only give the relevant part of a $\mathit{Neg}(\A)$ minimal model for this set of clauses: 
\begin{align*}
M=\lbrace &\ABox{} B(a), \ABox{} D(a), \ABox{} C(b), \ABox{} R(b,b), \ABox{} R(a,a),\\
	& \mathit{NegD}(a), \mathit{NegB}(a), \ldots\rbrace
\end{align*}
This model gives us the minimal deletion:
$\A' = \lbrace C(b), R(b,b), R(a,a) \rbrace$
\end{example}

Note that Theorem \ref{theorem:minmodel} can further be used for minimal deletion of a delete request $D$ which belongs to the deductive closure of the knowledge base but is not contained in the ABox $\A{}$. (Meaning $D \notin \A{}$ but $\T{}  \cup \A{} \models D$). In this case we only have to make sure, that $\sigma(D)\in S$. If $\sigma(D)$ does not occur in $\A{}$ we have to add $D$ manually to $S$ in order to render the instance deletion possible.

The $\mathcal{K}^{*}$-transformation introduced in Definition \ref{def:kstar} can be used to repair an ABox, which is inconsistent w.r.t. its TBox. The basic idea is to replace each occurrence of $\bot$ in $\T{}$ by a new atom $\mathit{false}$ and further add $\mathit{false}$ to $S$. After that, we use the $\mathcal{K}^{*}$-transformation and construct the minimal instance deletion of $\false$ from the ABox. The resulting ABox is a minimal ABox repair.

\begin{lemma}\label{lemma:minrepair}Let $\mathcal{K}=(\T,\A)$ be a knowledge base with consistent $\T{}$, $\T_{\false}$ the TBox obtained from $\T{}$ by replacing every occurrence of $\bot$ by $\false$, $\A_{\false}$ be $\A{} \cup \lbrace \false \rbrace$ and $\K_{\false} = (\T_{\false},\A_{\false})$. Let $S$ be the set of atomic concepts and roles occurring in $\A{}$ and $\T{}$ plus $\mathit{false}$. Then there is  a $\Neg(\A)$-minimal model for $\K_{\mathit{false}}^{*} \cup \lbrace \Negfalse\leftarrow \top \rbrace$. 
\end{lemma}

\begin{corollary} \label{corollary:aboxrepair}Let $\mathcal{K}=(\T,\A)$, $\T_{\false}$ and $S$ be defined as in Lemma \ref{lemma:minrepair}. Then $\A{} \setminus Del(M)$ is a minimal ABox repair for $\A{}$ for all $\Neg(\A)$-minimal models $M$ for $\K_{\mathit{false}}^{*} \cup \lbrace \Negfalse\leftarrow \top\rbrace$.
\end{corollary}
Corollary \ref{corollary:aboxrepair} follows immediately from Theorem \ref{theorem:minmodel} with $D=\mathit{false}$. See \cite{FurbachSchonTR2013} for both proofs.
Lemma \ref{lemma:minrepair} together with Corollary \ref{corollary:aboxrepair} implies, that such a minimal ABox repair can always be constructed.

Next we consider a special case of deletion.
For a given knowledge base $\mathcal{K}=(\T,\A)$, Theorem \ref{theorem:minmodel} can only be used to construct a minimal instance deletion of $D$ from $\A{}$ if $\K{}^ {*}\cup \lbrace \Neg(D) \leftarrow  \top  \rbrace$ is satisfiable. However if $\K{}^ {*}\cup \lbrace \Neg(D) \leftarrow \top \rbrace$  is not satisfiable, there is no $\Neg(\A)$-minimal model for $\K{}^ {*}\cup \lbrace \Neg(D) \leftarrow  \top \rbrace$  and therefore we cannot use Theorem \ref{theorem:minmodel} for the construction of a minimal instance deletion.
\begin{example}
Let $\T{}$ be a TBox containing the assertion 
$\top \sqsubseteq C$
stating that everything belongs to the concept $C$. This corresponds to the DL-clause 
$C(x)\leftarrow  \top$.
Let us further consider the ABox:
$\A=\lbrace C(a),B(a),C(b),B(b)\rbrace$
The $\mathcal{K}^{*}$-transformation leads to 
\begin{align*}
\K{}^ {*}=\lbrace    & C(x) \leftarrow \top, \nonumber \\
				& \bot \leftarrow \mathit{NegC}(x), \nonumber \\
				& \bot \leftarrow C(x) \land \mathit{NegC}(x), \nonumber \\
				& \mathit{ABoxC(a)},\nonumber \\
				& \mathit{ABoxB(a)},\nonumber \\
				& \mathit{ABoxC(b)},\nonumber \\
				& \mathit{ABoxB(b)}\rbrace\nonumber 
\end{align*}

If we now want to delete $C(a)$ from $\A{}$, we have to construct $\Neg(\A)$-minimal models for $\K{}^ {*} \cup \lbrace \mathit{NegC}(a) \leftarrow  \top \rbrace$. However $\K{}^ {*} \cup \lbrace \mathit{NegC}(a) \leftarrow  \top \rbrace$ is unsatisfiable. So we are not able to construct a minimal instance deletion of $C(a)$ from $\A{}$ using Theorem \ref{theorem:minmodel}.
Taking a closer look at the TBox reveals the problem: the TBox claims, that everything has to belong to the concept $C$. So the only way to remove $C(a)$ from $\A{}$ is to remove individual $a$ entirely from the ABox.
\end{example}

The next Theorem uses this idea and states how to construct minimal ABox deletions in the case that $\K{}^ {*} \cup \lbrace \mathit{Neg}(D) \leftarrow  \top  \rbrace$ is unsatisfiable. 
Please note that the requirement of $\T{}  \cup \A{}$ being consistent in the next theorem is not a limitation since we are always able to repair an ABox which is inconsistent with respect to its TBox using Corollary \ref{corollary:aboxrepair}.
\begin{theorem}\label{theorem:unsatdeletion} Let $\mathcal{K}=(\T,\A)$ be a knowledge base with $\T{}  \cup \A{}$ consistent. Let further $S$ be the set of atomic concepts and roles occurring in $\A{}$ and let $D$ be a delete request with $\mathit{Ind}(D)$ the set of individuals occurring in $D$.  If  $\K{}^ {*} \cup \lbrace \Neg(D)\leftarrow  \top \rbrace$ is unsatisfiable, then $\A' \subseteq \A{}$ is a minimal instance deletion of $D$ from $\A{}$, where $\A'$ is obtained from $\A{}$ by removing all ABox assertions containing an individual from $\mathit{Ind}(D)$.
\end{theorem}
With the help of Theorem \ref{theorem:minmodel} and \ref{theorem:unsatdeletion} we are now able to construct minimal instance-level deletions independent from the satisfiability of $\K{}^{*} \cup \lbrace \Neg(D)\leftarrow  \top \rbrace$.

%

Next we address the insertion of an assertion into an existing ABox. This can be obtained, by first adding the assertion to the ABox and afterwards constructing all possible minimal repairs for the resulting ABox. If the added assertion is not contained in any of these minimal ABox repairs, then it is not possible to insert the assertion into the ABox without rendering the ABox inconsistent w.r.t. its TBox. If there is a minimal repair containing the added assertion, then the insertion is possible and the respective minimal ABox repair gives us the result of the insertion.
\begin{example}In the Example \ref{ex:insertion}, we can repair $\A'$. There are two minimal ABox repairs for $\A'$: $\A''=\lbrace C(a) \rbrace$ and $\A'''=\lbrace D(a) \rbrace$. The first minimal repair corresponds to deleting the previously inserted $D(a)$ and therefore is not desirable. The second minimal repair however allows us to keep the inserted assertion.
\end{example}


\section{Experimental Results}\label{sect:experiments}
We developed a prototypical implementation for deletion of ABox assertions using the $\K{}^ {*}$-transformation. We use the E-KRHyper theorem prover to construct the $\Neg(\A)$-minimal models which lead us to the minimal deletions. Another theorem prover able to handle DL-clauses is HermiT \cite{msh07optimizing}. However HermiT is not able to calculate $\Neg(\A)$-minimal models. This is why we chose the E-KRHyper theorem prover for our implementation.
All tests were carried out on a computer featuring an AMD Phenom X6 1090T @ 3.2GHz and 8GB RAM.
To the best of our knowledge, there is no system performing deletion of ABox assertions as described in this paper. This is why we cannot compare our system to another system. 

%
 In Section \ref{sect:nf} we briefly discussed the complexity of the entire  $\K{}^ {*}$-transformation. There is a linear blow up of the knowledge base and there is also polynomial time complexity for performing the transformation. The real costs for performing the deletion, insertion and repair are caused by the theorem prover which has to compute the $\Neg(\A)$-minimal models. For an overview about this issue we refer to \cite{DBLP:books/el/RV01/DixFN01}.
We use E-KRHyper for the construction of $\Neg(\A)$-minimal models. For this we extended E-KRHyper by a feature to construct $\Gamma$-minimal models in a bottom-up way. This extension renders it possible to give E-KRHyper a set of DL-clauses together with a set of predicate symbols $P$ and an integer $i$. Then E-KRHyper only constructs models containing at most $i$ instances of $P$ predicates. During reasoning, E-KRHyper discards all models with more than $i$ instances of $P$ predicates. If E-KRHyper is not able to find a model with $i$ or less instances of $P$ predicates, it terminates by stating that the maximal number of instances is reached. We use this feature to construct $\Neg(\A)$-minimal models: for $S$ the set of concepts and roles occurring in the knowledge base, we first call E-KRHyper with $\K{}^ {*}$, the set $\Neg(S)$ and $i=1$. We successively increase $i$ until E-KRHyper either gives us a model or a proof for the unsatisfiability of the set of DL-clauses. This ensures that the first model given by E-KRHyper is a $\Neg(\A)$-minimal model.

\begin{figure}[t]
    \centering
    \begin{tikzpicture}
        \begin{axis}[
                width=\textwidth,
                height=0.5 \textwidth,
                font=\footnotesize,
                ymajorgrids=true,
                yminorgrids=true,
                scaled x ticks = false, 
                x tick label style={
                    /pgf/number format/.cd,
                    fixed,
                    fixed zerofill,
                    precision=0,
                    set thousands separator={}
                },
                xtick={3653,13653,23653,33653,43653,53653},
                ytick={5.0,4.5,4.0,3.5,3.0,2.5,2.0,1.5,1.0,0.5,0.0},
                xlabel style={yshift=-0.5em},
                xlabel=number of ABox assertions,
                ylabel style={yshift=-0.5em},
                ylabel=run time (sec),
                legend style={anchor=north east,legend pos=north west },
                no markers,
            ]
            \addplot[dashed] table[x=Size,y=atomic]{rt_ekrh.dat};
            \addplot[dotted] table[x=Size,y=non-atomic]{rt_ekrh.dat};
            \addplot[solid] table[x=Size,y=ekrh]{rt_ekrh.dat};
            \legend{$\K{}^{*}$-transformation (atomic),$\K{}^{*}$-transformation (non-atomic),E-KRHyper without $\K{}^ {*}$};
        \end{axis}
    \end{tikzpicture}
    \caption{Time used for atomic and non-atomic deletions.}
    \label{fig:numbersvicodi}
\end{figure}
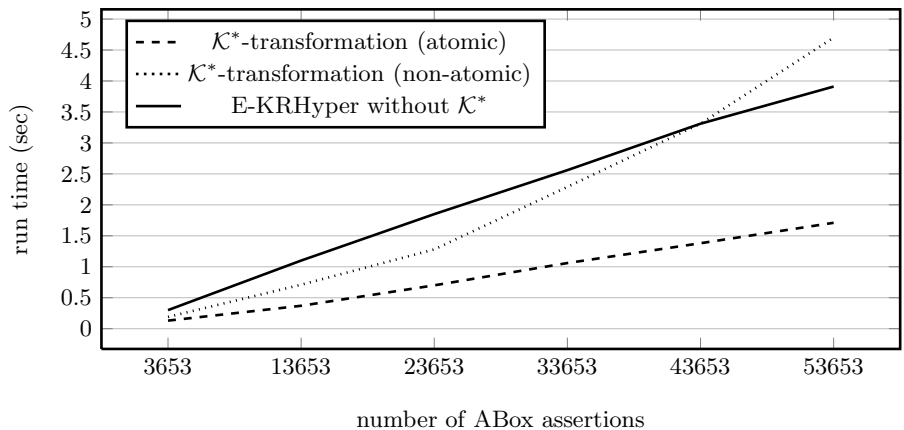

We use the $\mathcal{ALHI}$ ontology VICODI \footnote{http://www.vicodi.org} for testing our approach. The smallest version of this ontology consists of 223 axioms in the TBox and RBox and 53653 ABox assertions. The larger versions of this ontology are generated by duplicating the assertions of the original ABox several times and changing the names of  the individuals in the assertions. Unfortunately the repetitive structure of the larger versions of the ontology, resulting from this construction, is not suitable to test the efficiency of our approach. This is why we focus on the smallest version of the VICODI ontology.
We construct different versions with increasing numbers of ABox assertions. The TBox and RBox remain unchanged. 
%
 For each version of the so created ontologies we used 1000 different ABox assertions as a delete request $D$, calculated the $\K{}^{*}$-transformation and used E-KRHyper to calculate the minimal ABox deletion. In Figure \ref{fig:numbersvicodi} we show the results for the different ABox sizes we considered. 
For most of the delete request considered, it was sufficient to only remove the delete request itself from the ABox. We call those cases atomic deletions. If more than one ABox assertion has to be deleted, we speak of non-atomic deletions. 
Figure \ref{fig:numbersvicodi} gives information on the average time used for a delete request leading to an atomic deletion as well as leading to a non-atomic deletion. 
Another way to determine atomic deletions is to use E-KRHyper without the $\K{}^ {*}$-transformation. If we want to test, if $D$ can be removed from the ABox by deleting only $D$ from the ontology $KB$, we can test $KB \setminus \lbrace D \rbrace \cup \lbrace \lnot D \rbrace$ for satisfiability using E-KRHyper. Satisfiability of  $KB \setminus \lbrace D \rbrace \cup \lbrace \lnot D \rbrace$ implies, that $KB \setminus \lbrace D \rbrace \not\models D$. Meaning that $D$ can be deleted atomically.
Note that this test can only be used for atomic deletions and is completely useless for the calculation of non-atomic deletions.
You can find the time used for those atomic deletions computed by E-KRHyper without the $\K{}^ {*}$-Transformation in Figure \ref{fig:numbersvicodi}.
Comparing the lines for E-KRHyper and atomic deletions using the $\K{}^ {*}$-transformation shows, that the $\K{}^ {*}$-Transformation is faster in calculating atomic deletions. In addition to that the $\K{}^ {*}$-Transformation is able to calculate non-atomic deletions as well and is therefore better suited for deletion than E-KRHyper.
Figure \ref{fig:numbersvicodi} reveals another nice property of the $\K{}^{*}$-transformation: increasing the size of the ABox only leads to a harmless increase of the time necessary to calculate the minimal deletion. We owe this property to the fact, that we only calculate the deviation from the original ABox. For the calculation of non-atomic deletions more than one run of E-KRHyper is necessary. This explains why non-atomic deletions take longer than atomic deletions. However the time necessary to calculate a non-atomic deletion only increases moderately when the size of the ABox under consideration is increased.

\section{Conclusion and Future Work}
In this paper we give a semantically guided compilation technique, the so called $\K{}^ {*}$-transformation, for \shi{} knowledge bases. The transformed knowledge base is equisatisfiable to the original one. A theorem prover can be used for the computation of the necessary actions for  deletion, insertion and repair from the result of the $\K{}^{*}$-transformation. Especially theorem provers based on a hypertableau calculus are suited for these computations.
The approach is implemented and we introduced first experimental results using the theorem prover E-KRHyper.

In future work, we want to extend our implementation to enable it to do ABox repair and insertion of assertions as well. 

Since E-KRHyper is able to handle the DL \shiq{}, we plan to extend our approach to qualified number restrictions.


\vspace{1.2cm}

\bibliography{lit}
\end{document}